\documentclass[12pt]{iopart}
\usepackage{setstack}
\usepackage{iopams}
\newtheorem{fact}{Fact}

\newcommand{\ket}[1]{\mbox{$| #1 \rangle$}}
\newcommand{\bk}[2]{\ensuremath{\langle #1 | #2 \rangle}}

\newcommand{\Ad}{\mbox{Ad}}
\newcommand{\Stab}{\mbox{Stab}}

\begin{document}
\title[Geometry of the local equivalence of states]{Geometry of the local equivalence of states}
\author{A Sawicki$^{1,2}$ and M Ku\'s $^{1}$}

\address{$^1$Center for Theoretical Physics, Polish Academy of Sciences, Al.
Lotnik\'ow 32/46, 02-668 Warszawa, Poland}

\address{$^2$School of Mathematics, University of Bristol,
University Walk, Bristol BS8 1TW, UK} 

\eads{\mailto{assawi@cft.edu.pl},  \mailto{marek.kus@cft.edu.pl}}

\begin{abstract}

We present a description of locally equivalent states in terms of symplectic
geometry. Using the moment map between local orbits in the space of states
and coadjoint orbits of the local unitary group we reduce the problem of
local unitary equivalence to an easy part consisting of identifying the
proper coadjoint orbit and a harder problem of the geometry of fibers of the
moment map. We give a detailed analysis of the properties of orbits of
``equally entangled states''. In particular we show connections between
certain symplectic properties of orbits such as their isotropy and coisotropy
with effective criteria of local unitary equivalence.

\pacs{03.67.Mn, 03.65.Aa, 02.20.Sv}


\end{abstract}
\submitto{\JPA}
\maketitle

\section{Introduction}

In a recent paper \cite{sawicki11} we presented a symplectic description of
pure states of composite quantum systems in finite-dimensional Hilbert
spaces. In particular we showed that entanglement among subsystems of a
multipartite quantum system can be quantified in terms of degeneracy of the
canonical symplectic form on the complex projective space restricted to
orbits of local, i.e., entanglement preserving unitary groups. In the present
paper we would like to continue this line of research by giving a precise
geometric description of orbits in low dimensional cases and, above all, by
showing how the proposed geometric approach contributes to a solution of an
important problem of local unitary equivalence of states.

The classification of states which are connected by local unitary
transformations, i.e.\ operations on the whole system composed from unitary
actions (purely quantum evolutions) each of which is restricted to a single
subsystem has become recently a topic of several studies \cite{kraus10},
\cite{kraus10a}. To appreciate the experimental importance of such a
 setting let us remind that it is a basis for such spectacular
applications of quantum information technologies like teleportation
or dense codding where the fundamental parts of experiments consist
of manipulations restricted to parts of the whole system in distant
laboratories.

\section{Symplectic geometry of entanglement}
We start with a short outline of a symplectic description of quantum
correlations in composite systems. For details consult \cite{sawicki11}.
Thorough expositions of the below employed constructions from symplectic
geometry can be found in \cite{guillemin84} and \cite{kirillov04}.

\subsection{Space of quantum states as a symplectic manifold}
\label{subsec:space}

The Hilbert space of a quantum system consisting of $L$ identical
$N$-level systems (qunits) is the tensor product\footnote{A
generalization to nonidentical subsystems, i.e., living in spaces of
different dimensionality is straightforward but more tedious.},
\begin{equation}\label{Hqudit}
\mathcal{H}=\mathcal{H}_1\otimes\cdots\otimes\mathcal{H}_L,
\end{equation}
where each $\mathcal{H}_k$ is isomorphic with the complex $N$-dimensional
space $\mathbb{C}^N$ equipped with the standard Hermitian scalar product
$\bk{\cdot}{\,\cdot}$ (we will denote by the same symbol the standard scalar
product in the whole $\mathcal{H}$ as along as it does not lead to
confusion).

The set of pure states is the projective space $\mathbb{P}(\mathcal{H})$.
We denote a canonical projection from $\mathcal{H}$  to
$\mathbb{P}(\mathcal{H})$ by $\pi$ and use the notation $[v]=\pi(v)$ for
$v\in\mathcal{H}$.

The projective space $\mathbb{P}(\mathcal{H})$ is equipped with a natural
symplectic structure - the Fubini Study form - inherited from the initial
Hilbert space $\mathcal{H}$ where a natural symplectic structure is defined
in terms of the imaginary part of the scalar product. For further purposes it
is convenient to calculate the symplectic form on $\mathbb{P}(\mathcal{H})$
in the following way. First observe that the linear action of the unitary
group $U(\mathcal{H})$ on $\mathcal{H}$ projects in a natural way to
$\mathbb{P}(\mathcal{H})$ as
\begin{equation}\label{unitaryactionP}
U[v]:=[Uv], \quad U\in U(\mathcal{H}), \quad v\in\mathcal{H}, \quad [v]=\pi(v).
\end{equation}
Let $A\in\mathfrak{u}(\mathcal{H})=Lie(U(\mathcal{H}))$ (the Lie algebra of
$U(\mathcal{H})$ which also acts linearly on $\mathcal{H}$). Denote by
$T_{[v]}\mathbb{P}(\mathcal{H})$ the tangent space to
$\mathbb{P}(\mathcal{H})$ at the point $[v]$, and by $A_{[v]}$ the vector in
$T_{[v]}\mathbb{P}(\mathcal{H})$ tangent to the curve
$t\mapsto\pi(\exp(tA)v)$. When $A$ runs through the whole Lie algebra
$\mathfrak{u}(\mathcal{H})$ the corresponding $A_{[v]}$ span
$T_{[v]}\mathbb{P}(\mathcal{H})$ and the symplectic form on
$\mathbb{P}(\mathcal{H})$ at $[v]$ reads
\begin{equation}\label{sympP}
\fl\omega_{[v]}(A_{[v]},B_{[v]})=\mathrm{Im}\frac{\bk{Av}{Bv}\bk vv-
\bk{Av}v\bk v{Bv}}{\bk vv^{2}}=-\frac{i}{2}\frac{\bk{[A,B]v}{v}}{\bk{v}{v}},
\quad A,B\in\mathfrak{u}(\mathcal{H}),
\end{equation}
where $[\,\cdot\,,\cdot]$ is the Lie bracket (commutator) in
$\mathfrak{u}(\mathcal{H})$. One checks that indeed $\omega$ is nondegenerate
and closed, $d\omega=0$, on $T\mathbb{P}(\mathcal{H})$ and as such makes
$\mathbb{P}(\mathcal{H})$ a symplectic manifold. Moreover, as it is clear
from the above construction, $\omega$ is invariant with respect to the action
(\ref{unitaryactionP}) of $U(\mathcal{H})$. In other words the action of
$U(\mathcal{H})$ on $\mathbb{P}(\mathcal{H})$ is symplectic.

\subsection{Symplectic group actions. Moment map}\label{symplaction}

Symplectic actions of semisimple groups lead to another important
construction useful in our analysis - the moment map. Let a compact
semisimple group $G$ acts on a symplectic manifold $(M,\omega)$ \textit{via}
symplectomorhophisms $G\times M\ni(g,x)\mapsto\Phi_g(x)\in M$, i.e., we
demand that the pullback of the form $\omega$ by $\Phi_g$ is the form
$\omega$ itself, $\Phi_g^*\omega=\omega$. For an arbitrary
$\xi\in\mathfrak{g}=Lie(G)$ (the Lie algebra of $G$) we define a vector field
$\hat{\xi}$ (called in the following the fundamental vector field
corresponding to $\xi$),
\begin{equation}\label{fvf}
\hat\xi(x)=\frac{d}{dt}\bigg|_{t=0} \Phi_{\exp t\xi}(x).
\end{equation}
Since $G$ acts on $M$ by symplectomorhophisms there exists a function
$\mu_\xi:M\rightarrow\mathbb{R}$ such that
\begin{equation}\label{moment1}
d\mu_\xi=\imath_{\hat{\xi}}\,\omega:=\omega(\hat{\xi},\cdot).
\end{equation}
It can be chosen to be linear in $\xi$, i.e., there exists $\mu(x)$ in the
space of linear forms on $\mathfrak{g}$ (the dual space to $\mathfrak{g}$
denoted in the following by $\mathfrak{g}^\ast$) such that
\begin{equation}\label{moment-lin}
\mu_\xi(x)=\langle\mu(x),\xi\rangle,\quad
\mu(x)\in\mathfrak{g}^\ast,
\end{equation}
where $\langle\,,\rangle$ is the pairing between $\mathfrak{g}$ and
$\mathfrak{g}^\ast$. In this way we obtain a function
$\mu:M\rightarrow\mathfrak{g}^\ast$ called the moment map.

The group $G$ acts on its Lie algebra $\mathfrak{g}$ \textit{via} the adjoint
action,
\begin{equation}\label{ADg}
{\Ad}_g\xi=\frac{d}{dt}\bigg|_{t=0}g\,\exp t\xi\,
g^{-1}=:g\xi g^{-1},\quad g\in G, \quad \xi\in\mathfrak{g},
\end{equation}
which dualizes to the coadjoint action on $\mathfrak{g}^\ast$,
\begin{equation}\label{Adast}
\langle{\Ad}^\ast_g\alpha,\xi\rangle=\langle\alpha,{\Ad}_{g^{-1}}
\xi\rangle=\langle\alpha,g^{-1}\xi g\rangle,
\end{equation}
for $g\in G$, $\xi\in\mathfrak{g}$, and
$\alpha\in\mathfrak{g}^\ast$. Under our assumption of the
semisimplicity of $G$ the momentum map can be chosen equivariant,
i.e., for each $x\in M$ and $g\in G$,
\begin{equation}\label{moment-equiv}
\mu\left(\Phi_g(x)\right)={\Ad}^\ast_g \mu(x),
\end{equation}
is fulfilled.

Coadjoint orbits, i.e., the orbits of a coadjoint action of $G$ on
$\mathfrak{g}^*$ bear a canonical symplectic structure - the so called
Kirillov-Kostant-Souriau form. Let $\Omega_\alpha$ be the coadjoint orbit
going through $\alpha\in\mathfrak{g}^\ast$,
\begin{equation}\label{coadorbit}
\Omega_\alpha=\{{\Ad}^\ast_g\alpha:g\in G\}.
\end{equation}
For any $\xi\in\mathfrak{g}$ let $\tilde{\xi}$ be a vector tangent at
$\alpha$ to the curve $t\mapsto \Ad^\ast_{exp(t\xi)}\alpha$,
\begin{equation}\label{tangentvector}
\tilde{\xi}=\frac{d}{dt}\bigg|_{t=0}{\Ad}^\ast_{exp(t\xi)}\alpha.
\end{equation}
When $\xi$ runs over the whole algebra $\mathfrak{g}$ such vectors span the
tangent space to $\Omega_\alpha$ at the point $\alpha$. We define the desired
symplectic form $\tilde\omega$ at the point $\alpha$ by its action on two
vectors constructed \textit{via} (\ref{tangentvector}) from the $\xi$ and
$\eta$ elements of $\mathfrak{g}$,
\begin{equation}\label{symformcoa}
\tilde\omega_\alpha(\tilde{\xi},\tilde{\eta})=\langle\alpha,[\xi,\eta]\rangle.
\end{equation}
We can obviously repeat the construction at each point $\beta$ on
$\Omega_\alpha$ obtaining thus a symplectic form on the whole orbit. It can
be checked that $\tilde{\omega}$ constructed in this way is indeed closed and
nondegenerate on $\Omega_\alpha$, as well as $G$-invarint, i.e.,
$\left(Ad_g^\ast\right)^\ast\tilde\omega=\tilde\omega$.

Due to the equivariance of the moment map (\ref{moment-equiv}) the orbit of
the $G$-action on $M$ going through a point $x$,
\begin{equation}\label{orbit}
\mathcal{O}_x:=\{\Phi_g(x),\ g\in G\},
\end{equation}
is mapped by $\mu$ onto a coadjoint orbit,
\begin{equation}\label{coorbit}
\Omega_{\mu(x)}=\{\Ad_g^*\mu(x),\ g\in G\}.
\end{equation}
Moreover the map $\mu$ intertwines the symplectic structures on $M$ and
coadjoint orbits; if we pull back $\tilde\omega$ from $\Omega_{\mu(x)}$ by
$\mu$ to $M$ we recover the restriction of $\omega$ to $\mathcal{O}_x$,
\begin{equation}\label{intsympl}
\mu^*\tilde\omega=\omega|_{\mathcal{O}_x}.
\end{equation}
In this way we obtained a map between two symplectic structures
which can be used to investigate properties of $G$-orbits in $M$.
First natural questions which can be addressed with the help of the
above constructions concerns symplecticity of orbits. In general the
moment map does not map $\mathcal{O}_x$ onto $\Omega_{\mu(x)}$
diffeomorphically. If it were the case then all $G$-orbits in $M$
would be symplectic, i.e., the restriction of $\omega$ to an orbit
would be nondegenerate. To characterize fully the situation when it
is the case let us consider two subgroups of $G$ - the stabilizers
of, respectively, $x$ and $\mu(x)$,
\begin{eqnarray}
 \Stab(x) &=& \{g\in G: \Phi_g(x)=x\}, \label{stabx} \\
 \Stab(\mu(x)) &=& \{g\in G: \Ad_g^\ast\mu(x)=\mu(x)\}. \label{stabmux}
\end{eqnarray}
As a consequence of the equivariance of $\mu$ we have always
$\Stab(x)\subset\Stab(\mu(x))$. The Kostant-Sternberg theorem
\cite{kostant82} states that an $G$ orbit is symplectic if and only if both
stabilizers are equal. As a corollary we obtain that the degeneracy subspace
at $x$ defined as
\begin{equation}\label{degspace}
\mathcal{D}_{x}=\{u\in T_{x}\mathcal{O}_{x}:\omega|_{\mathcal{O}_{x}}(u,v)
=0\,\,\,\forall v\in T_{x}\mathcal{O}_{x}\}.
\end{equation}
has the dimension
\begin{equation}\label{dimdeg}
D(x)=\dim(\mathcal{D}_x)=\dim(\Stab(\mu(x)))-\dim(\Stab(x)),
\end{equation}
or, taking into account that $\dim(\Stab(x))=\dim(G)-\dim(\mathcal{O}_x)$ and
$\dim(\Stab(\mu(x)))=\dim(G)-\dim(\Omega_{\mu(x)})$,
\begin{equation}\label{dimdeg1}
D(x)=\dim(\mathcal{D}_x)=\dim(\mathcal{O}_x)-\dim(\Omega_{\mu(x)}).
\end{equation}
The dimension (\ref{dimdeg1}) is of course constant along the whole orbit.

\subsection{Orbits in the space of states. Entanglement}
For the entanglement problem of $L$ identical $N$-level subsystems the
relevant group $G$ is the $L$-fold direct product of the special unitary
group,
\begin{equation}\label{Gqunit}
G=SU(N)\times\cdots\times SU(N),
\end{equation}
acting in the natural way on the tensor product $\mathcal{H}$, i.e., $g\cdot
v= U_1v_1\otimes\cdots\otimes U_Lv_L$ for $g=(U_1,\ldots,U_L)\in G$,
$v=v_1\otimes\cdots\otimes v_L$, $v_k\in \mathcal{H}_k$. This action is
projected to the symplectic manifold
\begin{equation}\label{M}
M=\mathbb{P}(\mathcal{H}).
\end{equation}
$G$ is a group of local unitary transformations where each $SU(N)$ represents
unitary quantum operations exercised on a single subsystem placed in one
laboratory. They preserve quantum correlations among subsystems, i.e., they
leave the ``amount of entanglement'' in the system intact.

The moment map for the action of the unitary group $U(\mathcal{H})$
(isomorphic to $U(N^L)$) on $\mathbb{P}(\mathcal{H})$ is easily calculated as
\begin{equation}\label{momentPH}
\langle\mu([v]),A\rangle=\frac{i}{2}\frac{\bk{v}{Av}}{\bk{v}{v}},
\quad A\in\mathfrak{u}(\mathcal{H}).
\end{equation}
The group $G$ of local transformations (\ref{Gqunit}) is a subgroup of
$U(\mathcal{H})$. All relevant formulas for the  symplectic forms and the
moment map remain the same after appropriate restrictions to $G$ and its Lie
algebra $\mathfrak{g}=\mathfrak{su}(N)\oplus\cdots\oplus\mathfrak{su}(N)$.

In \cite{sawicki11} we showed that the only symplectic orbit in
$\mathbb{P}(\mathcal{H})$ is the manifold of separable (nonentangled) states
and the dimension\footnote{Here and in the following by $\dim$ we understand
always the real dimension of the corresponding linear spaces and manifolds,
even if they bear complex structures.} of degeneracy space
$\mathcal{D}_{[v]}$ can be used to quantify entanglement of a state $[v]$ or,
in other words, $D([v])$ is an entanglement measure.

\section{Local unitary equivalence of states}\label{LUE}

For simplicity, in the following, as it is customary, we will use the term
``states'' also for vectors from $\mathcal{H}$ remembering, however, that in
fact we have in mind their projections to $\mathbb{P}(\mathcal{H})$.

Two pure states $v,w\in\mathcal{H}$ are called locally unitary (LU)
equivalent if and only if there exist $U_i\in SU(N)$ such that
\begin{equation}\label{ludef}
[v]=U_1\otimes\cdots\otimes U_L[w],
\end{equation}
i.e., $[u]$ and $[w]$ belong to the same orbit of the action of $G$ on
$\mathbb{P}(\mathcal{H})$.

We will show how the above outlined symplectic description of entanglement
can help in analyzing local unitary equivalence. As a first step let us
calculate the image of an arbitrary state $[v]$ under the moment map
(\ref{momentPH}).

Choosing an orthonormal basis $\{e_k\}$, $k=1,\ldots, n$ in
$\mathcal{H}_i\simeq \mathbb{C}^N$ we can write an arbitrary
$v\in\mathcal{H}$ in the form
\begin{equation}\label{expabsis}
v=\sum\limits_{k_1,\ldots,k_L}C_{k_1\ldots k_L}e_{k_1}\otimes
\cdots\otimes e_{k_L}.
\end{equation}
Without losing generality we can assume that $v$ has the unit length.

The Lie algebra
$\mathfrak{g}=\mathfrak{su}(N)\oplus\cdots\oplus\mathfrak{su}(N)$ is spanned
by the matrices $X_1\otimes\cdots\otimes I$, $\ldots$, $I\otimes\cdots\otimes
X_k\otimes\cdots\otimes I$, $\ldots$, $I\otimes\cdots\otimes X_L$ with
antihermitian $X_k$. The dual $\mathfrak{g}^\ast$ can be identified with
$\mathfrak{g}$ \textit{via} the invariant bilinear form $(X,Y)=-\tr(XY)$,
$X,Y\in\mathfrak{g}$, i.e., $\mathfrak{g}^\ast\ni\alpha\sim X\in\mathfrak{g}$
if $\langle\alpha,Y\rangle=-\tr(XY)$ for an arbitrary $Y\in\mathfrak{g}$. For
convenience we supplement this identification by multiplication of $X$ by the
imaginary unit making elements of $\mathfrak{g}^\ast$ Hermitian. This is
irrelevant for the whole reasoning but allows to treat elements of
$\mathfrak{g}^\ast$ as physical observables. Let thus
\begin{equation}\label{gstarelement}
X=X_1\otimes\cdots\otimes I+\ldots +I\otimes\cdots\otimes
X_k\otimes\cdots\otimes I+\ldots +I\otimes\cdots\otimes X_L,
\end{equation}
be an element of $\mathfrak{g}^\ast$. A straightforward calculation gives
\begin{equation}\label{muXv}
\mu_X([v])=\sum_k^L\sum_{n,m}^NC^{(k)}_{mn}\bk{e_m}{X_ke_n},
\end{equation}
where
\begin{equation}\label{Cknm}
C^{(k)}_{nm}=\sum_{l_1,\ldots, l_{L-1}}
\overline{C}_{l_1\ldots n \ldots l_{L-1}}C_{l_1\ldots m \ldots l_{L-1}},
\end{equation}
where the overbar denotes the complex conjugation and the summation is over
all corresponding pairs of indices except those on the $k$-th places. The
positive semidefinite matrices $C^{(k)}$ are in fact the reduced density
matrices of the subsystems. In the following we will occasionally use the
notation $C^{(k)}([v])$ to exhibit explicitly the dependence of the reduced
density matrices on the original state $[v]$.

It is known (see e.g., \cite{kirillov04}) that each coadjoint orbit of a
compact group (such as our $G$) intersects the dual $\mathfrak{t}^\ast$ to
the maximal commutative subalgebra $\mathfrak{t}$ of $\mathfrak{g}$. In our
case $\mathfrak{t}^\ast$ is spanned by $I\otimes\cdots\otimes
Y_k\otimes\cdots\otimes I$ with diagonal Hermitian $Y_k$. In general a
coadjoint orbit intersects $\mathfrak{t}^\ast$ in several points connected by
elements of the Weyl group. By restricting to a particular Weyl chamber,
e.g., by demanding that the diagonal elements of the diagonal matrices $Y_k$
appear in the nonincreasing order, we get rid of this redundancy.

In \cite{sawicki11} it was shown that for a state $[u]$ its image under the
moment map $\mu([u])$ belongs to $\mathfrak{t}^\ast$ if and only if all
matrices $C^{(k)}([u])$ are diagonal,
$C^{(k)}([u])=\mathrm{diag}(p_{1k}^2,\ldots,p_{Nk}^2)$, and we have in this
case,
\begin{equation}\label{intersection}
\mu([u])=Y_1\otimes I\otimes\cdots\otimes I+\cdots
+I\otimes\cdots\otimes I\otimes Y_L,
\end{equation}
where, up to an irrelevant multiplicative constant,
\begin{equation}\label{Yk}
Y_k=
\mathrm{diag}\left(-\frac{1}{N}+p_{1k}^2,\ldots,-\frac{1}{N}+p_{Nk}^2\right).
\end{equation}

Since the orbit $\mathcal{O}_{[v]}$ is mapped by the moment map onto the
coadjoint orbit $\Omega_{\mu([v])}$ each $v\in\mathcal{H}$ can be transformed
by some group element $U_1\otimes\ldots\otimes U_L\in G$ to
$v^\prime\in\mathcal{H}$ such that the moment map $\mu([v^\prime])$ belongs
to $\mathfrak{t}^*$, i.e, the reduced density matrices $C^{(k)}([v^\prime])$
are diagonal, $C^{(k)}([v^\prime])=\mathrm{diag}(p_{1k}^2,\ldots,p_{Nk}^2)$,
$p_{1k}^2\ge\ldots\ge p_{Nk}^2$. The corresponding $U_k$ are recovered from
the matrices diagonalizing the reduced density matrices of the state $v$. If
$\tilde{U}_k^\dagger C^{(k)}([v])\tilde{U}_k=C^{(k)}([v^\prime])$ then
$U_k=\tilde{U}_k^T$ ($^T$ denotes the transposition). Such a state
$x^\prime=[v^\prime]$ is called in \cite{kraus10,kraus10a} the `sorted trace
decomposition' of the state $x=[v]$. Observe that in the case of two
subsystems ($L=2$), the transformation form $x$ to $x^\prime$ can be made
unique using the Schmidt decomposition of a bipartite state.

Let us now return to the question of the local unitary equivalence of two
states $x=[u]$ and $y=[w]$, $x,y\in\mathbb{P}(\mathcal{H})$. Obviously, the
equalities $C^{(k)}(x)=C^{(k)}(y)$, $k=1,\ldots,L$, give a necessary
condition for the local unitary equivalence of $x$ and $y$. If they are
fulfilled we can use matrices $\tilde{U}_k$ and $\tilde{V}_k$ diagonalizing,
respectively, $C^{(k)}(x)$ and $C^{(k)}(y)$ to transform $x$ and $y$ to their
sorted trace decompositions $x^\prime$ and $y^\prime$. The equality
$x^\prime=y^\prime$ is then a sufficient condition for the local equivalence
of $x=[v]$ and $y=[w]$ and, explicitly,
\begin{equation}\label{vUVw}
[v]=U_1^\dagger V_1\otimes\cdots\otimes U_L^\dagger V_L[w].
\end{equation}
It is also clear that if the spectra of reduced density matrices for $x$ and
$y$ are equal but $x^\prime\ne y^\prime$ the states $x$ and $y$ may still be
locally unitary equivalent. Indeed, the equality of spectra of the reduced
density matrices of $x$ and $y$ means that $\mu(x^\prime)=\mu(y^\prime)$. If
$y^\prime=\Phi_{g^\prime}(x^\prime)$ for some $g^\prime\in G$, which is
equivalent to $y=\Phi_g(x)$ for some $g\in G$, then due to the equivariance
of $\mu$,
\begin{equation}\label{equiv1}
\mu(x^\prime)=\mu(y^\prime)=\mu(\Phi_{g^\prime}(x^\prime))=
\Ad_{g^\prime}^{\ast}(\mu(x^\prime)),
\end{equation}
i.e., $g^\prime\in\Stab(\mu(x^\prime))$. Summarizing, $x=[v]$ and $y=[w]$
with the same spectra of reduced density matrices are equivalent if and only
if there exists $g^\prime\in \Stab(\mu(x^\prime))$ such that
$\Phi_{g^\prime}(x^\prime)=y^\prime$. Since
$\Stab(\mu(x^\prime))\supset\Stab(x^\prime)$ this can happen also for
$x^\prime\ne y^\prime$.

In a generic case when spectra of all reduced density matrices are
nondegenerate (there are no multiple eigenvalues) the stabilizer
$\Stab(\mu(x^\prime))$ consists of diagonal unitary matrices, as it is clear
from Eq.(\ref{Yk}) giving explicitly the value of the moment map at an
intersection with $\mathfrak{t}^\ast$. In this case the algorithm of deciding
the local unitary equivalence can be effectively applied. In nongeneric cases
$\Stab(\mu(x^\prime))$ is a subgroup of $G=SU(N)^{\otimes L}$. As long as it
is a proper subgroup the effort of checking local unitary equivalence can be
considerably eased, but if for all $k$ we have
$C^{(k)}(x^\prime)=\frac{1}{\sqrt{N}}I$ (`maximally entangled states') we
return to the full group $G$ since in this case
$\Stab(\mu(x^\prime))=SU(N)\otimes\cdots\otimes SU(N)$.

\section{Fibers of the moment map}\label{fibers}

From the preceding section it is clear that to make progress in
checking the local unitary equivalence we have to investigate closer
the fiber of the moment map at $x$ (we will omit $^\prime$ in the
following assuming that $x$ is already reduced to its sorted trace
form), i.e.
\begin{equation}\label{fiber}
\mathcal{F}_{x}:=\{z\in M:\mu(z)=\mu(x)\}
=\mu^{-1}(\mu(x)).
\end{equation}

Let $\{\xi_k\}$, $k=1,\ldots,d=\dim\mathfrak{g}$,  be a basis in the Lie
algebra $\mathfrak{g}$. The corresponding vector fields $\hat{\xi}_k$ at $x$
(c.f.\ Eq.~(\ref{fvf})) span the tangent space $T_x\mathcal{O}_x$ to the
orbit through $x$ at $x$. On the other hand, the fiber $\mathcal{F}_x$ is a
common level set of the functions $\mu_{\xi_k}$,
\begin{equation}\label{levelset}
\mathcal{F}_x=\{z\in M:\mu_{\xi_k}(z)=c_k\},\quad c_k=\mu_{\xi_k}(x), \quad
k=1,\ldots,d.
\end{equation}
Let us define:
\begin{equation}\label{kermu}
\mathrm{Ker}_x(d\mu):=\{a\in T_x M:d\mu_{\xi_k}(x)(a)=0,k=1,\ldots,d\}.
\end{equation}
From (\ref{moment1}) we have
\begin{equation}\label{dmuak}
d\mu_{\xi_k}(a)=\omega(\hat{\xi}_k,a),
\end{equation}
hence $a\in\mathrm{Ker}_x(d\mu)$ if and only if $a$ is $\omega$-orthogonal to
all $\hat{\xi}_k$ and since the latter span $T_x\mathcal{O}_x$ we obtain
\begin{equation}\label{f1}
\mathrm{Ker}_x(d\mu)=(T_x\mathcal{O}_x)^{\perp\omega}.
\end{equation}

Since (c.f. \ref{levelset}), $\mu_{\xi_k}$ are constant on $\mathcal{F}$, we
have $d\mu_{\xi_k}(a)=0$ for $a\in T_x\mathcal{F}_x$. Hence $T_x\mathcal{F}_x
\subset \mathrm{Ker}_x(d\mu)$ and finally
$T_x\mathcal{F}_x\subset(T_x\mathcal{O}_x)^{\perp\omega}$.

It is obvious that the above reasoning does not depend on the choice of a
particular point in $\mathcal{F}_x$, i.e.,
\begin{equation}\label{fact1}
T_y\mathcal{F}_x\subset(T_y\mathcal{O}_x)^{\perp\omega}, \quad y\in\mathcal{F}_x.
\end{equation}

A submanifold $P$ of a symplectic manifold $M$ is called coisotropic if for
arbitrary $y\in P$ we have $(T_yP)^{\perp\omega}\subset T_yP$. We conclude
thus that if $\mathcal{O}_x$ is coisotropic then
$\mathcal{F}_x\subset\mathcal{O}_x$. Indeed from (\ref{fact1}) and the
cosisotropy of $\mathcal{O}_x$ at each $y\in\mathcal{F}_x$ we have
$T_y\mathcal{F}_x\subset T_y\mathcal{O}_x$. Hence, in this case examining
whether some $y$ belongs to $\mathcal{O}_x$ (and, consequently whether $y$
and $x$ are LU-equivalent) reduces to checking if their sorted trace forms
$x^\prime$ and $y^\prime$ have the same image under the moment map.

The coisotropy of $\mathcal{O}_x$ is a sufficient but not necessary condition
for $\mathcal{F}_x\subset\mathcal{O}_x$, since even for a non-coisotropic
orbit the fiber can be fully contained in it.

Summarizing the reasonings presented in the preceding sections we may
formulate the following observations. Let us assume that for two states
$x=[v]$ and $y=[w]$ the necessary condition for LU equivalence is fulfilled,
i.e., the spectra of the reduced density matrices $C^k(x)$ and $C^k(y)$ are
equal for all $k=1,\ldots,L$ and let $x^\prime=[v^\prime]$ and
$y^\prime=[w^\prime]$ be the sorted trace forms of $x$ and $y$. Then,
\begin{enumerate}
\item If the spectra of $C^k(x)$ for all $k$ are non-degenerate then
    establishing the LU equivalence of $x$ and $y$ consist of checking
    whether tere exists a diagonal unitary $U$ such that
    $[v^\prime]=[Uw^\prime]$ which reduces to a straightforward
    calculation.
\item If some spectra of the reduced density matrices are degenerate the
    states are LU equivalent if the fiber of the moment map
    $\mathcal{F}_x$ is contained in the orbit $\mathcal{O}_x$. A
    sufficient but not necessary condition for such an inclusion is the
    coisotropy of the orbit $\mathcal{O}_x$.
\end{enumerate}

In the two-partite case ($L=2$) the LU equivalence is easily checked by
performing the Schmidt decomposition of both considered states. If the
non-zero Schmidt coefficients, equal to the square roots of the reduced
density matrices (in this case equal for both subsystems) are equal, the
states are LU equivalent. This simple criterion is reflected in the geometry
of orbits and fibers of the moment map, albeit not in the simplest possible
way consisting of the coisotropy of orbits. In the next section we give a
detailed analysis of $L=2$ case identifying coisotropic and non-coisotropic
orbits and showing that also the latter contain the whole corresponding
fibers of the moment map.

\section{Two-partite states}\label{sec:twopartite}

As already mentioned, for $L=2$ the reduction of a state $x=[v]$ to its
sorted trace form gives the Schmidt decomposition of $v$. We will assume that
this operation has been already performed, hence we assume that $v$ reads as
\begin{equation}\label{schmidt}
v=\sum_{k=1}^Np_ke_k\otimes f_k,
\end{equation}
where $\{e_k\}$ and $\{f_k\}$ are appropriate orthonormal bases in
$\mathbb{C}^N$. Let us denote by $m_0$ the number of vanishing $p_k$ and by
$m_l$ the multiplicity of the consecutive nonzero coefficients $p_k$, thus
$\sum_{l=0}^rm_l=N$, where $r$ is the number of different nonvanishing
coefficients in the Schmidt decomposition (\ref{schmidt}).

It was proved in \cite{sawicki11} (see also \cite{szk02}) that the dimensions
of orbits $\mathcal{O}_{[v]}$ and $\Omega_{\mu([v])}$ are given as,
\begin{eqnarray}
  \dim\left(\mathcal{O}_{[v]}\right)&=& 2N^2-2m_0^2-\sum_{n=1}^rm_n^2-1,
\label{dimo} \\
  \dim\left(\Omega_{\mu([v])}\right)&=& 2N^2-2\sum_{l=0}^rm_n^2.\label{dimomu}
\end{eqnarray}
From the above we can thus easily calculate the dimension of the
space $\omega$-orthogonal to $T_{[v]}\mathcal{O}_{[v]}$
\begin{equation}\label{dimort}
\fl\dim\left(\left(T_{[v]}\mathcal{O}_{[v]}\right)^{\perp\omega}\right)=
\dim(\mathbb{P}(\mathcal{H}))-\dim(\mathcal{O}_{[v]})
=(2N^2-2)-\dim(\mathcal{O}_{[v]})=2m_0^2-\sum_{n=1}^rm_n^2-1,
\end{equation}
and the dimension of the degeneracy space (see Eq.~(\ref{dimdeg1})),
\begin{equation}\label{dimdeg2}
D([v])=\dim(\mathcal{O}_{[v]})-\dim(\Omega_{\mu([v])})=\sum_{n=1}^r m_n^2-1.
\end{equation}
Observe that the degeneracy space (\ref{degspace}) consists of exactly those
vectors from $T_{[v]}\mathcal{O}_{[v]}$ which simultaneously belong to
$\left(T_{[v]}\mathcal{O}_{[v]}\right)^{\perp\omega}$ hence it is the part of
$\left(T_{[v]}\mathcal{O}_{[v]}\right)^{\perp\omega}$ contained in
$T_{[v]}\mathcal{O}_{[v]}$. Comparing (\ref{dimort}) and (\ref{dimdeg2}) we
infer that an orbit is coisotropic if and only if all coefficients in the
Schmidt decomposition (\ref{schmidt}) differ from zero. In this case, as we
showed above, fibers of the moment map are contained in the corresponding
orbits. We will prove that this is the case also for non-coisotropic orbits.

First, observe that we have the following direct sum decomposition of
subspaces
\begin{equation}\label{decomport}
\left(T_{[v]}\mathcal{O}_{[v]}\right)^{\perp\omega}=\mathcal{D}_{[v]}\oplus S,
\end{equation}
where $\mathcal{D}_{[v]}$ is the degeneracy space (\ref{degspace})
and $S$ is a symplectic subspace of dimension $2m_0^2$ spanned by
$e_k\otimes f_l$ and $ie_k\otimes f_l$, where $k$ and $l$ are such
that the corresponding $p_k$ and $p_l$ in (\ref{schmidt})
vanish\footnote{Remember that we treat $T_{[v]}M$ as a real vector
space, hence $e_k\otimes f_l$ and $ie_k\otimes f_l$ are different
vectors.}. The symplecticity of $S$ is obvious since $\omega$ is
nondegenerate on it. Checking that $S$ is indeed spanned by the
mentioned vectors is a matter of a short calculation. Let us notice
first that for any $e_k\otimes f_l$ and $ie_k\otimes f_l$ such that
$p_k=0=p_l$ in (\ref{schmidt}) we have
\begin{equation}\label{ort}
\bk{e_k\otimes f_l}{v}=0=\bk{ie_k\otimes f_l}{v},\quad
\bk{e_k\otimes f_l}{e_k\otimes f_l}=1=\bk{ie_k\otimes f_l}{ie_k\otimes f_l},
\end{equation}
which means that $e_k\otimes f_l$ and $ie_k\otimes f_l$ belong to $\in
T_{[v]}M$. On the other hand (see the remark above Eq.~(\ref{sympP}), each
element of $T_{[v]}M$ has the form $\xi_{[v]}$ with
$\xi=(A\otimes I+I\otimes
B)\in\mathfrak{g}=\mathfrak{su}(N)\otimes\mathfrak{su}(N)$. Using formula (\ref{sympP})
and (\ref{ort}) we see that
\begin{equation}\label{ort1}
\omega_{[v]}(e_k\otimes f_l,(A\otimes I+I\otimes B)_{[v]})=
\mathrm{Im}\bk{e_k\otimes f_l}{(A\otimes I+I\otimes B)v},
\end{equation}
for any $A,B\in\mathfrak{su}(N)$. Direct calculations give
\begin{equation}\label{ort2}
\bk{e_k\otimes f_l}{(A\otimes I+I\otimes B)v}=
\sum_{i=1}^Np_i(\bk{e_k}{Ae_i}\bk{f_l}{f_i}+\bk{e_k}{e_i}\bk{f_l}{Bf_i}).
\end{equation}
Notice that $\bk{e_k}{Ae_i}\bk{f_l}{f_i}+\bk{e_k}{e_i}\bk{f_l}{Bf_i}\neq 0$
if and only if $i=l$ or $i=k$ but then from our assumption $p_i=0$ which
means (\ref{ort2}) and (\ref{ort1}) vanish. Hence, $e_k\otimes f_l$ and
$ie_k\otimes f_l$ are elements of $S\subset
\left(T_{[v]}\mathcal{O}_{[v]}\right)^{\perp\omega}$. Comparing the
dimensions of $\dim S=2m_0^2$ with (\ref{dimort}) and (\ref{dimdeg2}) we
obtain (\ref{decomport}).

It is now enough to show that fibers of the moment map are not tangent to
$S$. Let us use again the notation $x=[v]=\pi(v)$ and assume the contrary,
i.e., that there exists a curve $t\mapsto x(t)\in\mathcal{F}$ with $x(0)=x$,
such that the tangent $\dot{x}(0)$ to it at $x$ belongs to $S$. Since in the
two-partite case the fiber is given as a level set of functions
$\mu_{I\otimes A_k}$, and $\mu_{A_k\otimes I}$, where $A_k$ span
$\mathfrak{su}(N)$ it follows that for an arbitrary $A\in\mathfrak{su}(N)$ we
have $\mu_{I\otimes A}(x(t))=\mathrm{const}$, $\mu_{A\otimes
I}(x(t))=\mathrm{const}$ and, consequently,
\begin{eqnarray}
\frac{d\mu_{I\otimes A}(x(t))}{dt}\Bigg|_{t=0}=0=
\frac{d\mu_{A\otimes I}(x(t))}{dt}\Bigg|_{t=0}, \label{firstder}\\
\frac{d^{2}\mu_{I\otimes A}(x(t))}{dt^{2}}\Bigg|_{t=0}=0=
\frac{d^{2}\mu_{A\otimes I}(x(t))}{dt^{2}}\Bigg|_{t=0} \label{secondder}.
\end{eqnarray}
The first condition is always fulfilled due to the definition (\ref{moment1})
of $\mu$,
\begin{equation}
d\mu_{I\otimes A}(\dot{x}(0))=\omega(\widehat{I\otimes A},\dot{x}(0))=0,
\end{equation}
since $\widehat{I\otimes A}$ belongs to $T_x\mathcal{O}_x$ and from the
assumption, $\dot{x}(0)\in S\subset
\left(T_x\mathcal{O}_x\right)^{\perp\omega}$. The condition (\ref{secondder})
reads explicitly
\begin{equation}\label{secondder1}
\dot{x}(0)^T\cdot[D^{2}\mu_{I\otimes A}]\cdot\dot{x}(0)
+D^{1}\mu_{I\otimes A}\cdot\ddot{x}(0)=0,
\end{equation}
where $D^1\mu_{I\otimes A}$ is the first derivative vector and
$[D^2\mu_{I\otimes A}]$ the second derivative matrix of the function
$\mu_{I\otimes A}$ at $x$. Since $\dot{x}(0)$ is tangent to $M$ at $[v]$
we have $\dot{x}(0)=B_{[v]}=[Bv]$ for some $B\in\mathfrak{su}(N^2)$ (see
again the remark above Eq.(\ref{sympP})), by a direct calculation we find
\begin{equation}\label{secondder2}
\dot{x}(0)^T\cdot[D^{2}\mu_{I\otimes A}]\cdot\dot{x}(0)=
-\frac{i\bk{[[I\otimes A\,,\,B\,]\,,\,B\,]v}{v}}{2\bk{v}{v}}.
\end{equation}

Our aim now is to show that there exist such an $A\in\mathfrak{su}(N)$ that
(\ref{secondder1}) is not fulfilled for any choice of $\dot{x}(0)\in S$. In
Appendix A we show that the goal is achieved by taking $A\in\mathfrak{su}(N)$
as a diagonal traceless matrix with $A_{kk}=i$ if $p_{k}\neq0$ and
$A_{kk}=-i\frac{N-m_0}{m_0}$ if $p_k=0$ in (\ref{schmidt}). Appendix B
contains a complete description of orbits in the simplest non-trivial example
of two qutrits ($N=2$, $L=3$).

\section{Geometric structure of orbits through GHZ states}\label{secGHZ}

As remarked at the end of Section \ref{LUE}, the method of checking the local
unitary equivalence based on comparison of the moment map images gives no
advantages when all reduced density matrices are proportional to the
identity. In this section we will show how it is reflected in the structure
of orbits through the so called Greenberger-Horn-Zeilinger (GHZ) states for
$L\ge 3$ qubits. The Hilbert space will be thus
$\mathcal{H}=(\mathbb{C}^2)^{\otimes L}$ with the real dimension $\dim
(\mathcal{H})=2^{L+1}$ so $\dim (\mathbb{P}(\mathcal{H}))=2^{L+1}-2$. We are
interested in orbits of the action of $G=SU(2)^{\times L}$ on
$\mathbb{P}(\mathcal{H})$. The Lie algebra $\mathfrak{g}$ of $G$ is spanned
by
\begin{eqnarray}
\mathcal{X}_k=iI\otimes\cdots\otimes\underset{k}{\underset{\wedge}{\sigma_x}}
\otimes\cdots\otimes I \\
\mathcal{Y}_k=iI\otimes\cdots\otimes\underset{k}{\underset{\wedge}{\sigma_y}}
\otimes\cdots\otimes I \\
\mathcal{Z}_k=iI\otimes\cdots\otimes\underset{k}{\underset{\wedge}{\sigma_z}}
\otimes\cdots\otimes I
\end{eqnarray}
where $\sigma_{x}$, $\sigma_{y}$, and $\sigma_{z}$ are the Pauli matrices and
$k=1,\ldots,L$. The fibers of the moment map are given as level sets of the
functions $\mu_{\mathcal{X}_k}$, $\mu_{\mathcal{Y}_k}$,
$\mu_{\mathcal{Z}_k}$, $k=1,\ldots,L$.

Let us consider the $L$-partite GHZ state
\begin{equation}
v_{L}=\frac{1}{\sqrt{2}}\left(\ket 0^{\otimes L}+\ket 1^{\otimes
L}\right),\label{eq:GHZ2}
\end{equation}
where in order to make the formulas more readable we switched to the
customary notation of the qubit states $e_1=\ket{0}$, $e_2=\ket{1}$
together with $\ket{k}\otimes\ket{l}=\ket{kl}$ and $\ket{kk\ldots
k}=\ket{k}^{\otimes L}$, etc. The matrices $C^{k}([v_L])$
(\ref{Cknm}) are the same for all $k$,
\begin{equation}
C^{k}([v_L])=\left(\begin{array}{cc}
\frac{1}{2} & 0\\
0 & \frac{1}{2}\end{array}\right).\label{eq:GHZ}
\end{equation}
For the GHZ states we have $\mu([v_L])=0$. Indeed,
\begin{equation}
\fl\langle\mu([v_{L}])\,,\,I\otimes I\otimes\ldots\otimes X_k\otimes
\ldots\otimes I\rangle= \frac{i}{2}\bk{v_{L}}{I\otimes
I\otimes\ldots\otimes X_k\otimes \ldots\otimes Iv_{L}}
=\frac{i}{4}\tr X_{k}=0,
\end{equation}
for an arbitrary $X_k\in\mathfrak{su}(2)$. Hence, for any two vectors
$A_{[v_L]},B_{[v_L]}\in T_{[v_{L}]}\mathcal{O}_{[v_{L}]}$, where
$A,B\in\mathfrak{g}$ we have
\begin{equation}\label{omegarestrict}
\omega(A_{[v_L]},B_{[v_L]})=-\frac{i}{2}\bk{[A\,,\,B]v_L}{v_L}=
\langle\mu([v_{L}])\,,\,[A\,,\,B]\rangle=0,
\end{equation}
since $\mu([v_L])=0$.

Notice that for any $L\geq 3$ (\ref{omegarestrict}) implies that
$T_{[v_{L}]}\mathcal{O}_{[v_{L}]}\subseteq\left(T_{[v_{L}]}\mathcal{O}_{[v_{L}]}\right)^{\perp\omega}$.
We will show now that the orbit $\mathcal{O}_{[v_{3}]}$ through $[v_{3}]$ is
Lagrangian\footnote{The Lagrangian subspace $U$ of symplectic space
$(V,\omega)$ is the minimally coisotropic ($U\subseteq U^{\perp\omega}$) and
at the same time maximally isotropic ($U\supseteq U^{\perp\omega}$) subspace
of $V$, i.e., for any coisotropic space $W\subset U$ we have $W=U$ and for
any isotropic space $W^\prime \supset U$ we have $W^\prime=V$. These
conditions imply that $U$ is Lagrangian if and only if $U= U^{\perp\omega}$,
hence $\omega|_U=0$ and $\dim U=\frac{1}{2}\dim V$. }, i.e.,
$T_{[v_{L}]}\mathcal{O}_{[v_{L}]}=\left(T_{[v_{L}]}\mathcal{O}_{[v_{L}]}\right)^{\perp\omega}$
whereas for $L\geq4$ it is isotropic, i.e.,
$T_{[v_{L}]}\mathcal{O}_{[v_{L}]}\subset\left(T_{[v_{L}]}\mathcal{O}_{[v_{L}]}\right)^{\perp\omega}$.
In other words for $L=3$ the orbit $\mathcal{O}_{[v_{L}]}$ is (minimally)
coisotropic and for  $L>3$ the orbit $\mathcal{O}_{[v_{L}]}$ is not
coisotropic.

The space $T_{[v_{L}]}\mathcal{O}_{[v_{L}]}$ is spanned by the
vectors
\begin{eqnarray}
\mathcal{X}_kv_L &=&
\frac{i}{\sqrt{2}}\left(\ket{0\ldots\underset{k}{\underset{\wedge}{1}}\ldots 0}
+\ket {1\ldots\underset{k}{\underset{\wedge}{0}}\ldots 1}\right), \label{gx} \\
\mathcal{Y}_kv_L &=&
\frac{1}{\sqrt{2}}\left(\ket{0\ldots\underset{k}{\underset{\wedge}{1}}\ldots 0}
-\ket {1\ldots\underset{k}{\underset{\wedge}{0}}\ldots 1}\right), \label{gy} \\
\mathcal{Z}_kv_L &=& \frac{i}{\sqrt{2}}\left(\ket 0^{\otimes L}-
\ket 1^{\otimes L}\right), \label{gz}
\end{eqnarray}
with $k=1,\ldots,L$. The above $2L+1$ vectors are mutually
orthogonal, hence

\begin{fact}The orbit $\mathcal{O}_{[v_{L}]}$ has dimension
\begin{equation}
\dim\mathcal{O}_{[v_{L}]}=2L+1.\label{eq:GZHdim}
\end{equation}
\end{fact}

Since $\frac{1}{2}\dim (\mathbb{P}(\mathcal{H}))=2^L-1$, we have as a
immediate consequence,
\begin{fact}The orbit $\mathcal{O}_{[v_{3}]}$
is Lagrangian, and hence the fiber of the moment map is contained inside
it. If $L\geq4$ then $\mathcal{O}_{[v_{L}]}$ is not Lagrangian.
\end{fact}

Indeed from (\ref{eq:GZHdim})
$\dim\mathcal{O}_{[v_3]}=7=2^3-1=\frac{1}{2}\dim (\mathbb{P}(\mathcal{H}))$,
whereas for $L\ge 4$ we have $2^{L}-1>2L+1$ hence the orbits have too small
dimension to be Lagrangian.

The fact that $\mathcal{O}_{[v_{3}]}$ is Lagrangian (so also coisotropic),
implies that necessary and sufficient condition for two states $[u]$ and
$[w]$ of three qubits to belong to $\mathcal{O}_{[v_{3}]}$ is
$\mu([u])=0=\mu([w])$.

For $L\geq4$ the fiber of the moment map is not entirely contained in
$\mathcal{O}_{[v_{L}]}$. We will show that in fact
$T_{[v_{L}]}\mathcal{F}_{[v_{L}]}=
(T_{[v_{L}]}\mathcal{O}_{[v_{L}]})^{\perp\omega}$.

Let $A_{[v_L]}\in T_{[v_{L}]}\mathcal{O}_{[v_{L}]}$, i.e,
$A_{[v_L]}=[Av_L]$ where $A$ is of the form (\ref{gx})-(\ref{gz}),
\begin{equation}\label{A}
A=iI\otimes\cdots\otimes\underset{k}{\underset{\wedge}{\sigma_\beta}}
\otimes\cdots\otimes I, \quad \beta\in\{x,y,z\}.
\end{equation}

The space $(T_{[v_{L}]}\mathcal{O}_{[v_{L}]})^{\perp\omega}$ is spanned by
these $B_{[v_L]}\in T_{[v_L]}\mathbb{P}(\mathcal{H})$ for which
$\omega_{[v_L]}(A_{[v_L]},B_{[v_L]})=0$. According to (\ref{sympP}) such
vectors $B_{[v_L]}$ have the form $[Bv_L]$ with
$B\in\mathfrak{u}(\mathcal{H})$ and
\begin{equation}\label{secondterm}
0=\bk{[A,B]v_L}{v_L}=-\bk{v_L}{[B,A]v_L}.
\end{equation}
We can choose
\begin{equation}\label{B}
B=i\sigma_{\alpha_1}\otimes\cdots\otimes\sigma_{\alpha_L},
\end{equation}
with $\alpha_i\in\{x,y,z,0\}$ and $\sigma_0=1$ since such vectors span
$\mathfrak{u}(\mathcal{H})$.

To prove that $T_{[v_{L}]}\mathcal{F}_{[v_{L}]}=
(T_{[v_{L}]}\mathcal{O}_{[v_{L}]})^{\perp\omega}$ we have to show that if
$B_{[v_L]}=[Bv_L]$ belongs to
$(T_{[v_{L}]}\mathcal{O}_{[v_{L}]})^{\perp\omega}$ then the curve $t\mapsto
[e^{itB}v_L]$ is contained in the fiber of the moment map, i.e.,
\begin{equation}\label{curveinfiber}
\langle\mu([v_L]), e^{-itB}Ae^{itB}\rangle=\bk{v_L}{\,e^{-itB}Ae^{itB}v_L}=0,
\end{equation}
for arbitrary $A$ and $B$ of the forms, respectively, (\ref{A}) and
(\ref{B}), fulfilling (\ref{secondterm}). To this end we employ the Hadamard
lemma,
\begin{equation}\label{hadam}
e^{-itB}Ae^{itB}=A+(-it)[B,A]+\frac{(-it)^2}{2!}[B,[B,A]]+
\frac{(-it)^3}{3!}[B,[B,[B,A]]]+\cdots .
\end{equation}
Now, using $\sigma_{\alpha_n}^2=I$ we have from (\ref{A}) and (\ref{B}),
\begin{equation}\label{doublecomm}
[B,[B,A]]=-iI\otimes\cdots\otimes[\sigma_{\alpha_k},[\sigma_{\alpha_k},
\sigma_\beta]]\otimes\cdots\otimes I.
\end{equation}
From the commutation relations for Pauli matrices
\begin{equation}\label{sigmacomm}
[\sigma_x,\sigma_y]=i\sigma_z, \quad [\sigma_y,\sigma_z]=i\sigma_x, \quad
[\sigma_z,\sigma_x]=i\sigma_y,
\end{equation}
we infer that the double commutator $[B,[B,A]]$ equals $A$ (possibly
up to the sign) or vanishes (if $\alpha_k=\beta$ or
$\sigma_{\alpha_k}=I$). Consequently in the expansion (\ref{hadam})
we encounter only the terms proportional to $A$ and $[B,A]$. But
$\bk{v_L}{Av_L}=\frac{1}{2}\tr \sigma_\beta=0$ and
$\bk{v_L}{[B,A]v_L}=0$ vanishes on the assumption
(\ref{secondterm}). This concludes a proof of
\begin{fact}The tangent space $T_{[v_L]}\mathcal{F}_{[v_L]}$
to the fiber of the moment map over $\mu([v_L])$ is exactly equal to
$\left(T_{[v_L]}\mathcal{O}_{[v_L]}\right)^{\perp\omega}$ and orbits
$\mathcal{O}_{[v_L]}$ are isotropic.
\end{fact}

\section{Multiqubit systems}

\noindent In this section using geometric properties of state $[v_{3}]$
described in previous section we present easy method of checking whether two
states $[u]$ and $[v]$ of three qubits are locally unitary equivalent. Notice
at the beginning that in case of two qubits states the necessary and
sufficient condition for this is given by equality of Schmidt decompositions.
For three qubits we already know that states for which $\mu([v])=0$ are
locally equivalent and lie on the orbit $\mathcal{O}_{[v_3]}$ which is
Lagrangian. For other states the following reasoning is crucial.

\noindent Let us consider the action of $\mathcal{G}=U(\mathcal{H})$ on the
complex projective space $\mathbb{P}(\mathcal{H})$. Let $x=[u]$ and $y=[v]$
be two points from $\mathbb{P}(\mathcal{H})$. Since $\mathcal{G}$ - action is
transitive on $\mathbb{P}(\mathcal{H})$ there is at least one unitary matrix
$U\in \mathcal{G}$ joining $x$ with $y$, i.e,
\begin{equation}
[Uu]=[v].\label{eq:relation-1}
\end{equation}
Let $V\neq U$ has the property (\ref{eq:relation-1}). Then,
\begin{equation}
Ux=y=Vx\Rightarrow U^{-1}V\in\Stab(x).
\end{equation}
Hence, there is $W\in\Stab(x)$ such that $V=UW$. It means that all matrices
joining  $x$ with $y$ are of the form $UW$ where $W\in\Stab(x)$. Let us
consider now three vectors $v^1$, $v^2$ and $v^3$ such that
\begin{equation}
\bk{v^{i}}{v^{i}}=1,\,\,\,\bk{v^{1}}{v^{2}}=0,\,\,\,\bk{v^{1}}{v^{3}}=0,\label{eq:rela}
\end{equation}
i.e., all $v^{i}$ are normalized to one and $v^{2}$, $v^{3}$ are orthogonal
to $v^{1}$. Notice that $v^{2}$ can be obtained from $v^{3}$ by action of
unitary matrix $U_{1}\in\Stab(v^1)$. Hence, the general form of the unitary
matrix joining  $v^{2}$ with $v^{3}$ is
\begin{equation}
U=U_{1}V\quad U_{1}\in\Stab(\ket{v^{1}}),\,
V\in\Stab(\ket{v^{2}}).\label{eq:relat2}
\end{equation}
In case of three qubits $\mathcal{H=}\mathbb{C}^{2}\otimes\mathbb{C}^{2}
\otimes\mathbb{C}^{2}$ and the group of interest is $G=SU(3)^{\times 3}$. The
direct consequence of property (\ref{eq:relat2}) is the following fact

\begin{fact}\label{Dwarownowaznestany}Two locally equivalent states
$[x]$ and $[y]$ are orthogonal to some state $[z]$ if and only if there exist
$U\in\Stab([z])\cap G$ such that $[Ux]=[y]$.
\end{fact}

\noindent Using this fact we will give a simple criterion to check the LU
equivalence of two states $x=[u]$ i $y=[v]$. Let us assume at the beginning
that $x$ and $y$ are already in the sorted trace form, i.e.,
\begin{equation}
\mu(x)=\mu(y)=X_{1}\otimes I\otimes I+I\otimes X_{2}\otimes I+I\otimes
I\otimes X_{3},\label{eq:mulu}\end{equation} where matrices $X_{i}$ are
diagonal and at least one of them, e.g., $X_{1}$ has nondegenerate spectrum.
Under this assumptions states $x$ and $y$ can be written in the form
\begin{eqnarray}
u=p_{11}\ket 0\otimes\ket{\Psi_{1}}+p_{12}\ket 1\otimes\ket{\Psi_{2}},
\nonumber \\
v=p_{11}\ket 0\otimes\ket{\Phi_{1}}+p_{12}\ket 1\otimes\ket{\Phi_{2}},
\label{eq:adf}
\end{eqnarray}
where $\bk{\Psi_{i}}{\Psi_{j}}=\delta_{ij}$ and
$\bk{\Phi_{i}}{\Phi_{j}}=\delta_{ij}$ ($\ket{\Psi_{i}}$ and $\ket{\Phi_{i}}$
are two-qubit states). From (\ref{eq:adf}) we see that necessary condition
for $x$ and $y$ to be locally equivalent is local equivalence of pairs
$\ket{\Psi_{1}}$, $\ket{\Phi_{1}}$ and $\ket{\Psi_{2}}$, $\ket{\Phi_{2}}$,
but this can be easily checked using Schmidt decomposition as these are
two-qubit states. Assume that necessary condition is fulfilled. Hence, there
exists a matrix $U_{2}\otimes U_{3}$ joining state $\ket{\Psi_{1}}$ with
$\ket{\Phi_{1}}$, i.e.,
\begin{equation}
v^{\prime}=U_{2}\otimes U_{3}u=p_{11}\ket 0\otimes\ket{\Phi_{1}}+
p_{12}\ket 1\otimes\ket{\Psi_{2}^{\prime}},
\end{equation}
where $\bk{\Phi_{1}}{\Psi_{2}^{\prime}}=0$ and
$\ket{\Psi_{2}^{\prime}}=U_{2}\otimes U_{3}\ket{\Psi_{2}}$. Notice that we
can still act on $v^{\prime}$ with $\Stab(\ket{\Phi_{1}})\cap K$. But from
Fact~\ref{Dwarownowaznestany}, using assumption that $\ket{\Phi_{2}}$ is
locally equivalent with $\ket{\Psi_{2}^{\prime}}$ and that both
$\ket{\Phi_{2}}$ and $\ket{\Psi_{2}^{\prime}}$ are orthogonal to
$\ket{\Phi_{1}}$ we obtain that $x$ is locally equivalent to $y$. Summing up,
states of three qubits (\ref{eq:adf}) are locally equivalent if and only if
the corresponding pairs of states of two qubits $\ket{\Psi_{1}}$,
$\ket{\Phi_{1}}$ and $\ket{\Psi_{2}}$, $\ket{\Phi_{2}}$ are locally
equivalent. Notice that this method can be used to investigate local
equivalence of states of four qubits, but only if at least one of the
matrices (\ref{Cknm}) $C^k$ has nondegenerate spectrum. The example of the
state for which all four matrices $C^k$ have degenerate spectrum is
$[v_{4}]$. In Section~\ref{secGHZ} we proved that the orbit
$\mathcal{O}_{[v_4]}$ is not lagrangian but isotropic and fiber of the moment
map  over $\mu([v_{4}])$ is not entirely contained inside the orbit
$\mathcal{O}_{[v_4]}$. In fact the dimension of the part which is not
contained in $\mathcal{O}_{[v_4]}$ is $12$ and this makes the problem of
local equivalence hard.

\section{Summary}
The presented symplectic approach to entanglement exhibited
\textit{a priori} unexpected geometric richness of the space of pure
states for multipartite, finite dimensional quantum systems and shed
some light on the important problem of the local unitary equivalence
of pure states, or in physical terms, possibility of transforming
one state into another by means of quantum operations restricted to
single parties.

Using a fundamental concept of symplectic geometry and symplectic
group action theory, \textit{viz.} the moment map, the problem of
the local equivalence of states is mapped from the space of states
and corresponding orbits of local unitary groups onto geometry of
(co)adjoint orbits in corresponding local Lie algebras and their
duals. The procedure has an obvious advantage - checking whether two
elements of the Lie algebra or its dual space belong to the same
orbit (i.e. are ``locally equivalent'') reduces to the comparison of
spectra of (anti)symmetric matrices. On the other hand since the
moment map usually is not a diffeomorphism of an orbit in the space
of states onto the corresponding coadjoint orbit a detailed
investigation of its fiber is needed for the ultimate check of the
local equivalence of states. Such an analysis also clearly
identifies situations in which a conclusive solution is hard to
find.

The simplest situation occurs when an orbit of the local action in the space
of states is coisotropic. In this case the whole fiber is included in the
orbit and checking whether a state belongs to the orbit and hence is locally
equivalent to all other states on it consists of checking if the spectra of
all reduced density matrices are the same as for any other state on the
orbit. However, such a situation typically occurs only in various
``nondegenerate'' cases. On the other hand fibers can be fully contained in
the corresponding orbits also when the latter are not coisotropic. We have
illustrated such phenomena by analyzing the bipartite case. For two particles
checking of the local unitary equivalence of states can be effectively and
easily done by comparison of the Schmidt spectra. This fact should be
reflected in a simple geometry of local orbits. Indeed, we have shown that
only when no Schmidt coefficient vanishes the orbit is cosotropic;
nevertheless also non-coisotropic orbits contain the whole corresponding
fibers.

In order to interpret geometrically the principal obstacles for
effective checking the local unitary equivalence we analyzed the
local orbits through multiqubit GHZ states. For such states all
reduced density matrices are proportional to the identity (the
``maximally mixed'' states). The geometry of orbits through the GHZ
states depends on the number of parties. For three qubits the orbit
is Lagrangian, hence coisotropic. Consequently, the fiber of the
moment map is contained in it which means that all states that have
maximally mixed density matrices are locally unitary equivalent to
the GHZ state. We also showed that if the number of qubits exceeds
three the orbits through the GHZ states are isotropic rather than
coisotropic, and the corresponding fibers are only partially
included in them. This is the main obstacle for an easy effective
checking of the local unitary equivalence.

We believe that our approach to quantum entanglement discription,
although involving relatively abstract concept of symplectic
geometry, has already proven to be fruitful. It not only gives an
insight into geometric foundations of quantum mechanics but also
contributes to the solutions of important problems of quantum
information theory, hence the further continuation of this line of
research seems to be very promising.

\section*{Acknowledgments}

We gratefully acknowledge supports from the Polish Ministry of Science and
Higher Education through the project no. N N202 090239 and the Deutsche
Forschungsgemeischaft through the grant SFB-TR12.

\appendix
\setcounter{section}{1}
\section*{Appendix A}\label{AppendixA}

We will fill some details of the calculations showing that in the two-partite
case fibers of the moment map lie within the corresponding orbit. In
particular we will show that the fibers are not tangent to $S$ (see
(\ref{decomport})).

Let us define following operators
\begin{equation}
X_{ij}=i(E_{ij}-E_{ji}),\quad Y_{ij}=E_{ij}+E_{ji},\quad
H_{ij}=E_{ii}-E_{jj},\quad i<j,
\end{equation}
where $E_{ij}$ are matrices defined as
\begin{equation}
(E_{ij})_{kl}=\cases{
0 & for $k\neq i$,\, $l\neq j$,\\
1 & for $k=i$ i,\, $l=j$.}
\end{equation}
Without losing generality we assume that $p_{1}\neq0$ in (\ref{schmidt}).
Notice that vectors from $S$ can be generated in the following way
\begin{equation}\label{genS}
ip_{1}^{2}e_k \otimes f_l=[iY_{1k}\otimes Y_{1l}v],\quad
p_{1}^{2}e_k \otimes
f_l=[iY_{1k}\otimes X_{1l}v]
\end{equation}
Let us choose $A\in\mathfrak{su}(N)$ as a diagonal traceless matrix with
$A_{kk}=i$ if $p_{k}\neq0$ and $A_{kk}=-i\frac{N-m_0}{m_0}$ if $p_k=0$ in
(\ref{schmidt}). We have
\begin{equation}
D^{1}\mu_{I\otimes A}(\ddot{x}(0))=\omega(\widehat{I\otimes
A}\,,\,\ddot{x}(0))= \omega(0\,,\,\ddot{x}(0))=0.
\end{equation}
We used $\widehat{I\otimes A}=[I\otimes Av]$ which follows from the fact that
$I\otimes Av=iv$ and as such it corresponds to the zero vector in the tangent
space $T_{[v]}M$.  What is left to be shown is thus
\begin{equation}
\dot{x}(0)^T[D^{2}\mu_{I\otimes A}]\dot{x}(0)\neq0.\label{eq:ost}
\end{equation}
for any $\dot{x}(0)\in S$. Let us thus write
\begin{equation}
\dot{x}(0)=\sum_{k,l}(a_{kl}e_k\otimes f_l+b_{kl}ie_k\otimes f_l)
=p_1^{-2}\sum_{k,l}[i(a_{kl}Y_{1k}\otimes X_{1l}+b_{kl}Y_{1k}\otimes Y_{1l})v]
\end{equation}
where the sum goes over such $k,l$ that $p_{k}=0=p_{l}$ in (\ref{schmidt})
and we used (\ref{genS}) to obtain the second equality.

To calculate explicitly the second derivative using (\ref{secondder2}) we
need some commutators,
\begin{eqnarray}
[I\otimes A\,,\, iY_{1k}\otimes Y_{1l}]=iY_{1k}\otimes[A,\, Y_{1l}]=i\alpha
Y_{1k}\otimes X_{1l},\\{} [I\otimes A\,,\, iY_{1k}\otimes
X_{1l}]=iY_{1k}\otimes[A,\, X_{1l}]=-i\alpha Y_{1k}\otimes Y_{1l},
\end{eqnarray}
where $\alpha=\frac{N}{m_0}$. Hence,
\begin{equation}
[I\otimes A\,,\,\sum_{k,l}i(a_{kl}Y_{1k}\otimes X_{1l}+b_{kl}Y_{1k}\otimes Y_{1l})]=i\alpha\sum_{k,l}b_{kl}Y_{1k}\otimes X_{1l}-a_{kl}Y_{1k}\otimes Y_{1l}.
\end{equation}
And, finally,
\begin{eqnarray}
\fl\dot{x}(0)[D^{2}\mu_{I\otimes A}]\dot{x}(0)=\\\nonumber
\fl=-i\bk{[[I\otimes A\,,\,p_1^{-2}\sum_{k,l}i(a_{kl}Y_{1k}\otimes X_{1l}+b_{kl}Y_{1k}\otimes Y_{1l})]\,,\,p_1^{-2}\sum_{k,l}i(a_{kl}Y_{1k}\otimes X_{1l}+b_{kl}Y_{1k}\otimes Y_{1l})]v}{v}=\\\nonumber
\fl=-i\bk{[i\alpha p_1^{-2}\sum_{k,l}(b_{kl}Y_{1k}\otimes X_{1l}-a_{kl}Y_{1k}\otimes Y_{1l})\,,\, ip_1^{-2}\sum_{k,l}(a_{kl}Y_{1k}\otimes X_{1l}+b_{kl}Y_{1k}\otimes Y_{1l})]v}{v}=\\\nonumber
\fl=-i\alpha \cdot\omega(\sum_{kl}b_{kl}e_k\otimes f_l-a_{kl}ie_k\otimes
f_l\,,\,\sum_{kl}a_{kl}e_k\otimes f_l+b_{kl}ie_k\otimes f_l)=-2i\alpha
\sum_{kl}(a_{kl}^{2}+b_{kl}^{2}).\end{eqnarray} This clearly means
that $\dot{x}(0)[D^{2}\mu_{I\otimes A}]\dot{x}(0)\neq 0$ for any
$\dot{x}(0)\in S$ and proves that in the bipartite case fibers of the moment
map are fully contained in the corresponding orbits.

\section*{Appendix B. Two Qutrits}
In case of two qutrits ($N=3$, $L=2$) the Hilbert space is
$\mathcal{H}=\mathbb{C}^{3}\otimes\mathbb{C}^{3}$ and $\dim
(\mathcal{H})=18$, so $\dim (\mathbb{P}(\mathcal{H}))=16$. The Lie algebra
$\mathfrak{g}=\mathfrak{su}(3) \oplus \mathfrak{su}(3)$ of $G=SU(3)\times
SU(3)$ is spanned by $\{A_k\otimes I\,,I\otimes A_k\}$, where
$\{A_k,k=1,\ldots,8\}$ is a basis in $\mathfrak{su}(3)$ hence $\dim
(\mathfrak{g})=16$. The fibers of moment map through $v$ are given as common
level set of sixteen functions $\mu_{A_k\otimes I}$, $\mu_{I\otimes A_k}$.
Without loosing generality we assume that the bases $\{e_k\}$ and $\{f_k\}$
in both Hilbert spaces are equal. As previously we switch to the customary
notation $e_1=\ket{0}=f_1$, $e_2=\ket{1}=f_2$, $e_3=\ket{2}=f_3$, together
with $\ket{kl}=\ket{k}\otimes\ket{l}$.

The general form of a Schmidt-decomposed two-qutrit state is given by
\begin{equation}
v=p_{1}\ket{00}+p_{2}\ket{11}+p_{3}\ket{22},
\label{qutrit}
\end{equation}
where $p_{1}^{2}+p_{2}^{2}+p_{3}^{2}=1$. There are six cases to consider.

\subsection*{1. {${p_{1}=1,\,\,\, p_{2}=p_{3}=0}$ (a separable state) }}
In this case $v=\ket{00}$. The orbit $\mathcal{O}_{[v]}$ through $[v]$ is
symplectic (\cite{sawicki11}) hence the part of the fiber which is contained
in $\mathcal{O}_{[v]}$ is zero dimensional. Orthogonal complement
$(T_{[v]}\mathcal{O}_{[v]})^{\perp\omega}$ is spanned by
\begin{equation}
\{\ket{22},\, i\ket{22},\,\ket{11},\, i\ket{11},\,\ket{12},\,
i\ket{12},\,\ket{21},\, i\ket{21}\},
\end{equation}
and is a symplectic vector space $S$. The matrix $A\in \mathfrak{su}(3)$ used
in the proof in Appendix A has the form
\begin{displaymath}
A=\left(\begin{array}{ccc}
  i & 0 & 0 \\
  0 & -\frac{i}{2} & 0 \\
  0 & 0 & -\frac{i}{2} \\
\end{array}\right).
\end{displaymath}
There is no fiber and the orbit is not coisotropic.

\subsection*{2. {$p_{1}=p_{2}=p_{3}=\frac{1}{\sqrt{3}}$ (the maximally
entangled state)}}

In this case the orbit  $\mathcal{O}_{[v]}$ through
$v=\frac{1}{\sqrt{3}}(\ket{00}+\ket{11}+\ket{22})$ is coisotropic since all
$p_k\neq 0$. In fact $\mathcal{O}_{[v]}$ is minimally coisotropic hence
Lagrangian, i.e.,
\begin{eqnarray}
(T_{[v]}\mathcal{O}_{[v]})^{\perp\omega} =T_{[v]}\mathcal{O}_{[v]},\\
\dim (T_{[v]}\mathcal{O}_{[v]})= \frac{1}{2}\dim \mathbb{P}(\mathcal{H}).
\end{eqnarray}
Using formula (\ref{dimdeg2}) it is easy to prove that in case of two qunits
it is always true that orbit through
\begin{equation}\label{maxentsate}
v=\sum_{k=1}^{N}\frac{1}{\sqrt{N}}\ket{kk},
\end{equation}
is Lagrangian. Namely for (\ref{maxentsate}) we have
\begin{equation}
(T_{[v]}\mathcal{O}_{[v]})^{\perp\omega}=D([v])=N^2-1= \frac{1}{2}\dim
\mathbb{P}(\mathcal{H}),
\end{equation}
hence $\mathcal{O}_{[v]}$ is Lagrangian \cite{bengtsson07}.
\subsection*{3. {$p_{1}\neq p_{2}\neq p_{3}\neq0$ (a generic state)}}
\noindent The orbit  $\mathcal{O}_{[v]}$ through
$v=p_{1}\ket{00}+p_{2}\ket{11}+p_{3}\ket{22}$ is coisotropic since all
$p_k\neq 0$. Formulas (\ref{dimo}) and (\ref{dimomu}) give
\begin{equation}
\dim (\mathcal{O}_{[v]})=14,\,\,
\dim (\mu(\mathcal{O}_{[v]}))=12.
\end{equation}
The whole is fiber is contained in $\mathcal{O}_{[v]}$ and is
two-dimensional.
\subsection*{4. {$p_{1}=p_{2}\neq0,\,\,\, p_{3}\neq0$}}
\noindent The orbit  $\mathcal{O}_{[v]}$ through
$v=p_{1}(\ket{00}+\ket{11})+p_{3}\ket{22}$ is coisotropic since all $p_k\neq
0$. Formulas (\ref{dimo}) and (\ref{dimomu}) give
\begin{equation}
\dim (\mathcal{O}_{[v]})=12,\,\,\dim (\mu(\mathcal{O}_{[v]}))=8
\end{equation}
The whole fiber is contained in $\mathcal{O}_{[v]}$ and is four-dimensional.
\subsection*{5. {$p_{1}=p_{2}=\frac{1}{\sqrt{2}},\,\,\, p_{3}=0$}}
\noindent The orbit  $\mathcal{O}_{[v]}$ through
$v=\frac{1}{\sqrt{2}}(\ket{00}+\ket{11})$ is not coisotropic since $p_3=0$.
Formulas (\ref{dimo}) and (\ref{dimomu}) give
\begin{equation}
\dim (\mathcal{O}_{[v]})=11,\,\,
\dim (\mu(\mathcal{O}_{[v]}))=8.
\end{equation}
Hence the part of the fiber contained in $\mathcal{O}_{[v]}$ is
three-dimensional. The orthogonal complement
$(T_{[v]}\mathcal{O}_{[v]})^{\perp\omega}$ is five-dimensional and is spanned
by three vectors contained in $T_{[v]}\mathcal{O}_{[v]}$ and two other
$\{v_{1}=\ket{22},\,v_{2}=i\ket{22}\}$. The matrix $A\in \mathfrak{su}(3)$
used in the proof in Appendix A has the form
\begin{displaymath}
A=\left(\begin{array}{ccc}
  i & 0 & 0 \\
  0 & i & 0 \\
  0 & 0 & -2i \\
\end{array}\right).
\end{displaymath}  The whole fiber is contained inside
the orbit although the orbit is not coisotropic.
\subsection*{6. $p_{1}\neq p_{2}\neq0,\,\,\, p_{3}=0$}
\noindent The orbit  $\mathcal{O}_{[v]}$ through
$v=p_{1}\ket{00}+p_{2}\ket{11}$ is not coisotropic since $p_3=0$. Formulas
(\ref{dimo}) and (\ref{dimomu}) give
\begin{equation}
\dim (\mathcal{O}_{[v]})=13,\,\,\dim (\mu(\mathcal{O}_{[v]}))=12
\end{equation}
Hence the part of the fiber contained in $\mathcal{O}_{[v]}$ is
one-dimensional. The orthogonal complement
$(T_{[v]}\mathcal{O}_{[v]})^{\perp\omega}$ is three-dimensional and is
spanned by one vector contained in $T_{[v]}\mathcal{O}_{[v]}$ and two other
$\{v_{1}=\ket{22},\,v_{2}=i\ket{22}\}$. The matrix $A\in \mathfrak{su}(3)$
used in the proof in Appendix A has the form
\begin{displaymath}
A=\left(\begin{array}{ccc}
  i & 0 & 0 \\
  0 & i & 0 \\
  0 & 0 & -2i \\
\end{array}\right).
\end{displaymath} Again the whole fiber is contained inside the orbit although the orbit is not
coisotropic.

\section*{References}

\end{document}